\documentclass[usenatbib]{mn2e}

\usepackage{graphicx}
\usepackage{amssymb}

\title[GRO J1008--57]{Now you see it, now you don't - the circumstellar
disk in the GRO J1008--57 system}

\author[M.J. Coe et al.]{M. J.~Coe$^{1}$, A.J. ~Bird$^{1}$, A.B. ~Hill$^{1}$,
V.A. ~McBride$^{1}$, M. ~Schurch$^{1}$  \and J. Galache$^{1,2}$,
C. A. ~Wilson$^{3}$, M. ~Finger$^{3}$, D.A. ~Buckley$^{4}$ \& 
E. Romero-Colmenero$^{4}$\\
$^{1}$ School of Physics and Astronomy, Southampton University, SO17 
1BJ, UK\\ 
$^{2}$ Harvard-Smithsonian Center for Astrophysics, 60 Garden Street,
Cambridge, MA 02138, USA. \\
$^{3}$ NASA/MSFC, Huntsville, AL 35812, USA\\
$^{4}$ South African Astronomical Observatory, P.O. Box 9,
Observatory, 7935, South Africa.}

\begin{document}

\date{16 Apr 2007}

\pagerange{\pageref{firstpage}--\pageref{lastpage}} \pubyear{2002}

\maketitle

\label{firstpage}

\begin{abstract}

Multiwavelength observations are reported here of the Be/X-ray binary
pulsar system GRO J1008--57. Over ten years worth of data are gathered
together to show that the periodic X-ray outbursts are dependant on
both the binary motion and the size of the circumstellar disk. In the
first instance an accurate orbital solution is determined from pulse
periods, and in the second case the strength and shape of the
H$\alpha$ emission line is shown to be a valuable indicator of disk
size and its behaviour. Furthermore, the shape of the emission line
permits a direct determination of the disk size which is in good
agreement with theoretical estimates. A detailed study of the pulse
period variations during outbursts determined the binary period to be
247.8$\pm$0.4d, in good agreement with the period determined from the
recurrence of the outbursts.

\end{abstract}


\section{Introduction and background}

The Be/X-ray systems represent the largest sub-class of massive X-ray
binaries.  A survey of the literature reveals that of the 115
identified massive X-ray binary pulsar systems (identified here means
exhibiting a coherent X-ray pulse period), most of the systems fall
within this Be counterpart class of binary.  The orbit of the Be star
and the compact object, presumably a neutron star, is generally wide
and eccentric.  X-ray outbursts are normally associated with the
passage of the neutron star close to the circumstellar disk (Okazaki
\& Negueruela 2001). A review of such systems may
be found in Coe et al. (2000).

The source that is the subject of this paper,
the X-ray transient GRO J1008--57, was discovered on July 14 1993 by the
BATSE experiment on the Compton Gamma Ray Observatory (Stollberg et al. 
1993 and Wilson et al.  1994).  The source was observed in the 20--200
keV band, and was found to be pulsating at a period of 93.5s.  Its
spectrum was observed to be consistent with optically thin thermal
bremsstrahlung with kT=25 keV.  From these X-ray data it was concluded
that the system was a massive binary system - a neutron star with either
a Be or supergiant primary. A full report of this discovery outburst was
presented in Shrader et al. (1999).

Subsequently the optical counterpart was identified by Coe et al, (1994)
who showed it to be a V=15.3 OB star with a strong IR excess and strong
Balmer line emission. In this paper we present more than 10 years of
optical monitoring of this counterpart, including a 
blue-end spectrum taken from the Southern African Large
Telescope (SALT), which enables refinement of the spectral class to
much higher accuracy than previous published work. We also demonstrate
how the optical characteristics correlate with the long-term X-ray
(RXTE/ASM data) behaviour. This is supplemented by $\gamma$-ray
observations from INTEGRAL, in particular, reporting the details of
the June 2004 outburst from GRO J1008--57.

\section{Optical data}

\subsection{Red spectra}

H$\alpha$ data have been collected over the last 13 years from a
series of telescopes. The dates and properties of the H$\alpha$ line
are presented in Table~\ref{tab:obs}. In this table the following
telescopes and configurations have been used:

\begin{itemize}

\item AAT - 3.9m telescope, AAT (Australia), Royal Greenwhich
Observatory Spectrograph (RGOS), 25cm camera, 1200V grating and
TEK (1024x1024) CCD. The dispersion was 0.8\AA/pixel and the signal to
noise ratio from the 600s exposure was $\sim$40.

\item SAAO - 1.9m telescope, Sutherland Observatory (South Africa),
spectrograph, SITe detector, 1200 l/mm grating. The dispersion was
1.0\AA/pixel and the signal to noise ratio $\sim$10.

\end{itemize}

\begin{table}
\caption{Table of H$\alpha$ measurements. See text for details of
observatory/instrument used.}
\begin{tabular}{ccll}
\hline
Observation & Observatory & H$\alpha$ EW & Peak\\
Date & & (\AA)  & Separation\\
 & & & (km~s$^{-1}$)\\
\hline
21 Dec 1993	& SAAO & 	-17$\pm$2 & \\
24 Dec 1993	& SAAO &	-13.6$\pm$1.3 & 250 \\
26 Feb 1994	& AAT &		-19.4$\pm$1.2 & 215 \\
07 Mar 1994	& SAAO &	-19.6$\pm$0.5 & \\
26 Feb 1995	& SAAO &	-19$\pm$3 & \\
19 Nov 1995	& SAAO &	-11$\pm$2 & \\
20 Jun 1997	& SAAO &	-2.6$\pm$1.5 & \\
20 Jun 1997	& SAAO &	-4.9$\pm$1.3 & 350\\
06 Feb 1998	& SAAO &	-6.7$\pm$1.2 & \\ 
11 Jan 1999	& SAAO &	-19.0$\pm$1.6 & \\
14 Dec 2002	& SAAO &	-20$\pm$2 & \\
15 Mar 2005	& SAAO &	-22.6$\pm$1.7 & \\
19 Mar 2005	& SAAO &	-21.5$\pm$1.5 & 243 \\
07 May 2006	& SAAO &	-20.9$\pm$1.2 & \\
\hline

\end{tabular}
\label{tab:obs}
\end{table}

\begin{figure}
\begin{center}
\includegraphics[width=80mm,angle=-0]{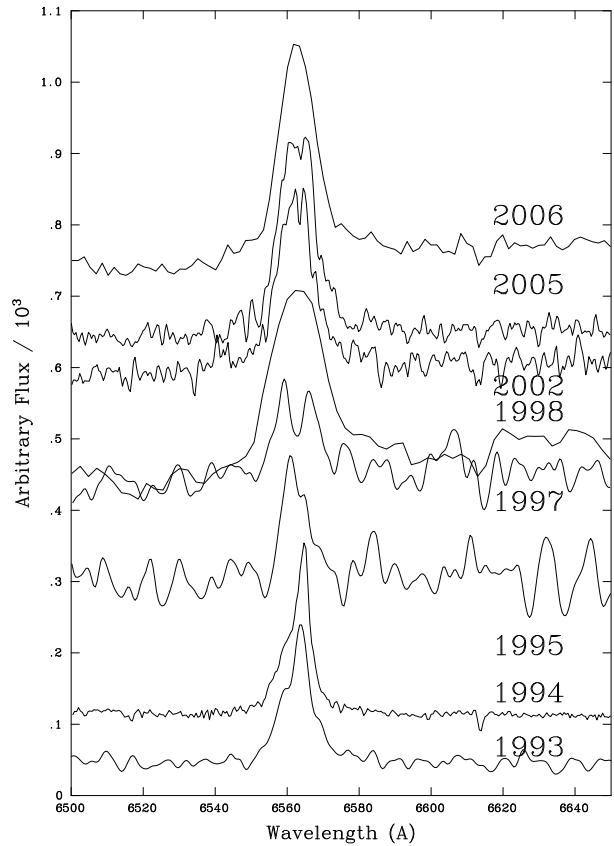}
\caption{The evolution of the H$\alpha$ profile over 13
years. Individual spectra have been arbitrarily re-scaled in flux to permit
the shape of the H$\alpha$ profile to be clearly visible.}
\label{fig:ha}
\end{center}
\end{figure}

\begin{figure}\begin{center}
\includegraphics[width=65mm,angle=90]{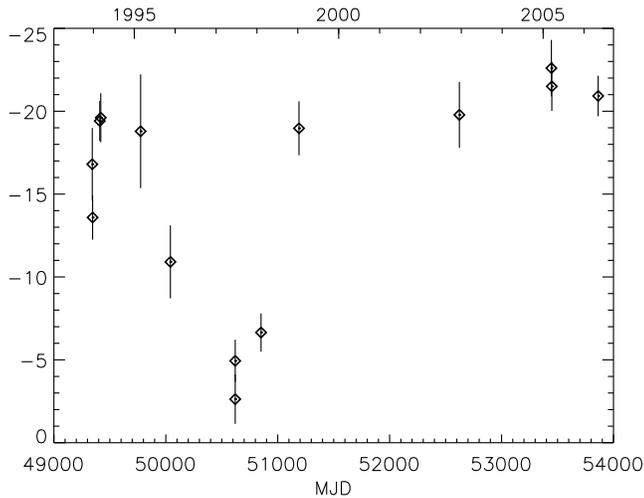}
\caption{The evolution of the flux of the H$\alpha$ emission line over 13 years
(1993--2006).}
\label{fig:ew2}
\end{center}
\end{figure}

Selected H$\alpha$ line profiles are presented in Figure~\ref{fig:ha},
where more than one spectra exists for a particular year then an
average of the two is shown. A
clear evolution of the Balmer line over the 10 year period is
seen. Apart from the structural changes which will be discussed below,
the strength of the H$\alpha$ emission line varies considerably over
the $\sim$10 year period - see Figure~\ref{fig:ew2} . The equivalent
width determinations can be seen to go through a significant decline
in the middle of the period, recovering strongly in the latest data.

The peak separations were determined by fitting Lorentzian profiles to
the H$\alpha$ emission lines.  In the cases where the $\chi^{2}$ was
significantly improved by including two lines, as opposed to one line,
the line parameters were used to measure the separation (in velocity
space) between the red and violet peaks. These kind of changes in the
line profiles are common in Be stars (see, for example, Telting \&
Kaper 1994).

\subsection{Blue spectrum}

\begin{figure}
\begin{center}
\includegraphics[width=70mm,angle=90]{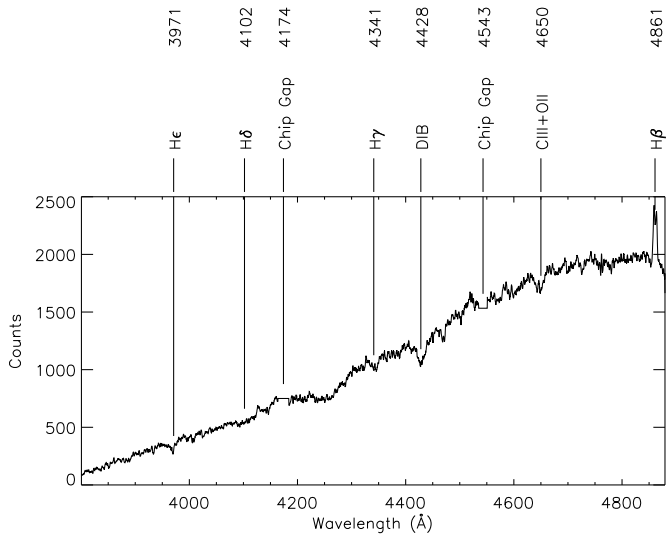}
\caption{SALT blue spectrum of the optical counterpart to GRO J1008--57. Major features in the spectrum are indicated.}
\label{fig:salt}
\end{center}
\end{figure}

A blue end spectrum was obtained from SALT on 18 June 2006 - see
Figure~\ref{fig:salt}. The spectrum was obtained using the Robert
Stobie Spectrograph in long slit mode, with a grating resolution of
1800 l/mm and an exposure time of 600s. This gave a dispersion of
0.34\AA/pixel and a signal to noise varying from $\sim$10 (blue end)
to $\sim$50 (red end).

Mosaicking of the three chips was performed by using the SALT IRAF
packages (the two chip gaps are clearly visible in
Figure~\ref{fig:salt}) as part of the pipeline processing of the
data. Further wavelength calibrations were performed using an Ar arc
lamp solution.  No flux calibration has been made.

The prominent feature is the strong double peaked H$\beta$ line, which
has an equivalent width of $-1.74\pm0.09$\AA.  The only previous blue
end spectrum (Coe et al.1994) gives a H$\beta$ equivalent width of
$-1.4\pm0.1$\AA. Two other Balmer lines (H$\gamma$ and H$\epsilon$)
are visible; but H$\delta$ seems to have been effectively lost due to
infilling.  This is unusual given that the Balmer lines on either side
do not seem to be so affected. The spectrum has been carefully checked
for any evidence of scattered light or other instrument atrtefacts,
but no evidence for any such features were found. The lack of any
identifiable HeII lines, in particular 4686\AA\, indicates that the
star is B1 or later (Walborn and Fitzpatrick 1990).  Furthermore the
presence of the OII CIII blend at 4650\AA\ restricts the
classification to the range B1-B2 (Walborn and Fitzpatrick 1990). It
is worth noting, however, that the lack of other metal lines is a
little surprising but may be due, in part, to the decreasing
signal-to-noise ratio towards the blue end of the spectrum. However,
such a lack of lines is not uncommon in low metalicity environments
such as the Small Magellanic Cloud (McBride et al, 2007).

\section{Orbital X-ray profile}

X-ray monitoring data were collected by the All Sky Monitor (ASM)
instrument on the Rossi X-ray Timing Explorer (RXTE) observatory. The
data consist of daily flux averages covering the X-ray energy range
1.3 -- 12.1 keV. There is no evidence in the daily data averages for
any significant detection of the source on short timescales ($\sim$ a
few days).

The $\sim$10 years worth of data were folded at the proposed
binary period of 248.9d determined by Levine \& Corbet, 2006 from the
same data. A strong, clear modulation at that period is very obvious.

\begin{figure}\begin{center}
\includegraphics[width=85mm,angle=-0]{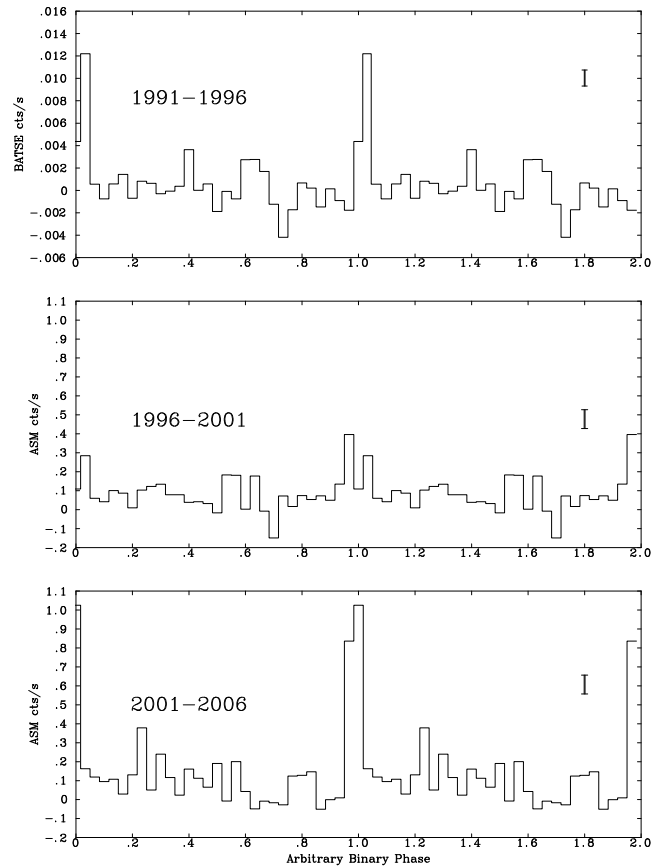}
\caption{BATSE (top panel) and RXTE/ASM (lower two panels) 
data folded at the period
of 248.9d (Levine \& Corbet, 2006) for three different time
intervals. The phases shown are those corresponding to the ephemeris
given in Section 6.1. In each case a $\pm1\sigma$ error bar is shown.}
\label{fig:fold}
\end{center}
\end{figure}

Since it is clear from Figure~\ref{fig:ew2} that the strength of the
H$\alpha$ line has been significantly changing over the 10 year interval
of the ASM data, the complete data set was divided into two equal
periods (1996-2001 \& 2001-2006). Each data set was separately folded
at this same period and with the same phase. The results are shown in
the lower two panels of Figure~\ref{fig:fold} from which it is clear
that the degree of modulation had changed very significantly over the
10 years. 

To explore the modulation at even earlier times (i.e. before 1996)
data from the BATSE occultation instrument on Compton Gamma Ray
Observatory (CGRO) were folded at the same period and same
phase. These data cover the harder spectral range of 20 -- 100 keV,
but reveal a very similar modulation occurring in the period 1991-1996
- see upper panel in Figure~\ref{fig:fold}. The phase of the outburst
and the degree of modulation looks very similar to that seen with
RXTE/ASM in recent years.

\section{CGRO/BATSE orbital solution}

The data types that are available from the 8 BATSE detectors on CGRO are the
16-channel continuous or ``CONT'' data, sampled at 2.048s intervals,
and the 4-channel discriminator or ``DISCLA'' data sampled every
1.024s. For the analyses in the present work the DISCLA data were used
from which pulsed fluxes and frequencies were extracted (see Bildsten
et al.1997 for a full explanation on the use of DISCLA data for pulsar
observations).

To assess the confidence in the detection of a known pulsar, the data
were searched through in a narrow frequency range centred on the
pulsar's nominal frequency.

First the DISCLA rates were grouped into ~300 s intervals. A pulse
profile was then estimated using a 6th order Fourier expansion for
each of these intervals. These profiles were then grouped into 4-day
intervals. Within each 4-day interval, the Fourier coefficients for
each 300-s pulse profile were re-computed for a grid of offsets from
the nominal frequency. The best frequency offset was determined using
the $Z_n^2$ statistic (Buccheri 1983), expressed as:

\begin{equation}
Z_n^2 = \sum_{k=1}^{N} \frac{|\mu_{k_{min}}|^2}{\sigma^2_{\mu_k}}
\end{equation}

\noindent where $\mu_{k_{min}}$ are the mean Fourier coefficients 
and $\sigma_{\mu_k}$ their Poisson errors; $N$ can take any value from
1 to 6 (in general $N$ = 3 or 6 was used). 
In addition, within each 4-day interval, a mean pulse profile was
determined for each frequency offset by chi-square
minimization. Unfortunately, BATSE noise is generally non-Poissonian
(Bildsten et al. 1997), so $\sigma_{\mu_k}$  is multiplied by the reduced
chi-square factor to produce the new statistic $Y_n$.
Monte-Carlo methods are required to
convert $Y_n$ into percentile significances in order to establish
confidence levels for the detected signal. An advantage to using this
method is that any number of harmonics can be excluded from the
statistic if there are other sources within the field of view with
pulsations at those particular harmonic frequencies.

If an outburst lasts for a significant fraction of a system's binary
orbital period, the size and shape of the orbit can be determined by
analysing Doppler shifts and pulse arrival delays in the pulsar
signal. As an outburst progresses, BATSE obtains a continuous
history of the daily source intensity and frequency; this detected
frequency will be the intrinsic spin frequency of the pulsar plus or
minus a shift due to the neutron star's orbit around its
companion. The process of calculating the orbit is further complicated
by the fact that the neutron star is usually spinning up or down
during an outburst, and so it is necessary to dissentangle the orbital
from the spin torque effects.

The method used with the BATSE data is the piecewise approach
explained in detail in Wilson et al, 2003. In essence, it assumes that
the torque during an outburst will be approximately constant during
small time intervals, such that the intrinsic spin up will be linear
during this time. As such, the data available during an outburst will
be divided into short segments of 3--5 data points each.  In order to
compute the orbital parameters of the system, the data in all the
segments are fitted with global values of $P_{orb}$, $\tau_{p}$,
{\it{e}}, $\omega$ and $a_{x}sin{i}$, but a different $\nu_{intr}$ in
each segment.

\begin{figure}
\begin{center}
\includegraphics[width=90mm,angle=0]{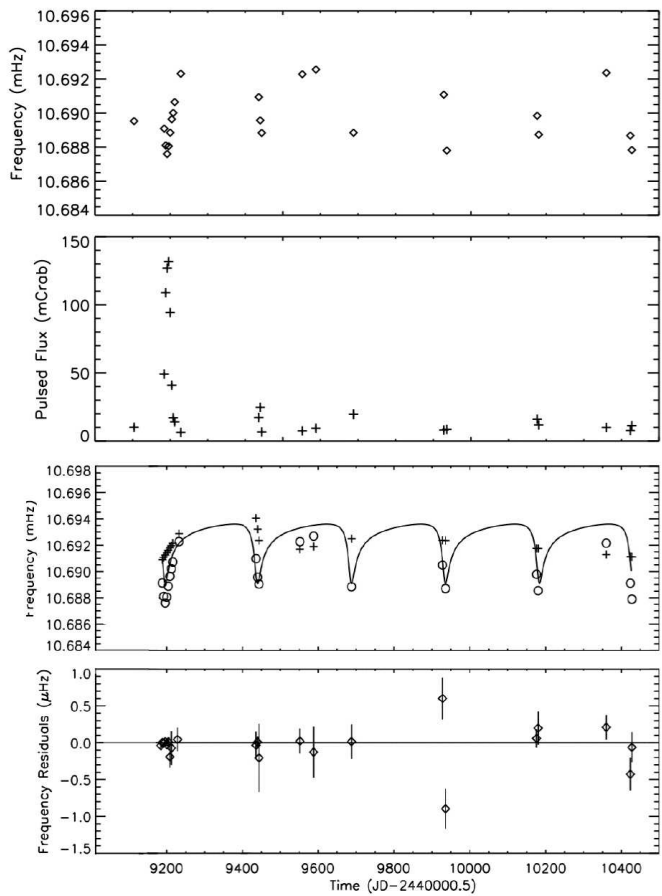}
\caption{BATSE data of GRO J1008--57. Top panel shows 4-day averaged 
frequencies and the second panel down shows the 
pulsed flux (n.b. only the $Y_n$$\ge$99.9\% significance
data points are shown). 
Third panel: the crosses are the emitted
frequency returned by fitting the model described in the text; the
solid line illustrates the magnitude of the orbital effect and the
circles are the frequencies as measured by BATSE. Lowest panel : pulse
frequency residuals for the model. 
}
\label{fig:combined}
\end{center}
\end{figure}

GRO J1008--57 was first detected during a bright outburst in July 1993
by BATSE with pulsations at 93.587$\pm$0.005s (Stollberg et al,
1993). This outburst can be seen around JD 2449200 in the second from
top panel in Figure~\ref{fig:combined}.  After this initial outburst,
a number of less significant detections were made, which could
correspond to 5 further outbursts. These are spaced $\sim$260d apart,
which led to this being proposed as the orbital period of the system
(Finger et al.1994). Shrader et al. (1999) find weak evidence for a
$\sim$135d period in the ASM data from RXTE.

Orbit fitting calculations were carried out using the $\ge$99.9\%
significance data points (all the points illustrated in
Figure~\ref{fig:combined}), producing an orbital period of
247.8$\pm$0.4d. This is in good agreement with the Levine \& Corbet
(2006) value of 248.9$\pm$0.5d. The results can be seen graphically in
Figure~\ref{fig:combined} and the orbital parameters are presented in
Table~\ref{table_j1008_orbit_params}.

Using the frequencies corrected for orbital motion from
Figure~\ref{fig:combined} we derive for the initial outburst a
$\dot{P} = -{5.6}\times 10^{-9}$ss$^{-1}$ implying L$_{x} = {1.0}
\times 10^{36}$erg s$^{-1}$ (B = ${1.4} \times 10^{13}$G).  Compare to
L$_{x} = {4.1} \times 10^{36}$ergs$^{-1}$ (B = ${6.0} \times
10^{13}$G), obtained by Shrader et al. 1999.

\begin{table}
  \begin{center}
  \caption{Orbital parameters for GRO~J1008-57.}
  \begin{tabular}[c]{l c}
  \noalign{\smallskip}
  \noalign{\smallskip} \noalign{\smallskip}
  $P_{orbital}$ (d)          &  247.8$\pm$0.4   \\
  $e$                            &  0.68$\pm$0.02   \\
  $a_{x}sini$ (light-s)             &  530$\pm$60      \\
  $\tau_{periastron}$ (MJD)  &  49189.8$\pm$0.5 \\
  \noalign{\smallskip}
  \end{tabular}
  \label{table_j1008_orbit_params}
  \end{center}
\end{table}

\section{INTEGRAL $\gamma$-ray data}

GRO J1008--57 was detected by INTEGRAL on two occasions - June 2004 and
October 2005 (Grebenev et al., 2005 have reported on the latter
outburst). On many other occasions the source has fallen within the
INTEGRAL field of view.The data from the IBIS imaging instrument
on-board INTEGRAL were reduced and analysed using the standard
\textit{Offline Science Analysis, OSA}, software version 5.1.

\begin{figure}\begin{center}
\includegraphics[width=70mm,angle=-90]{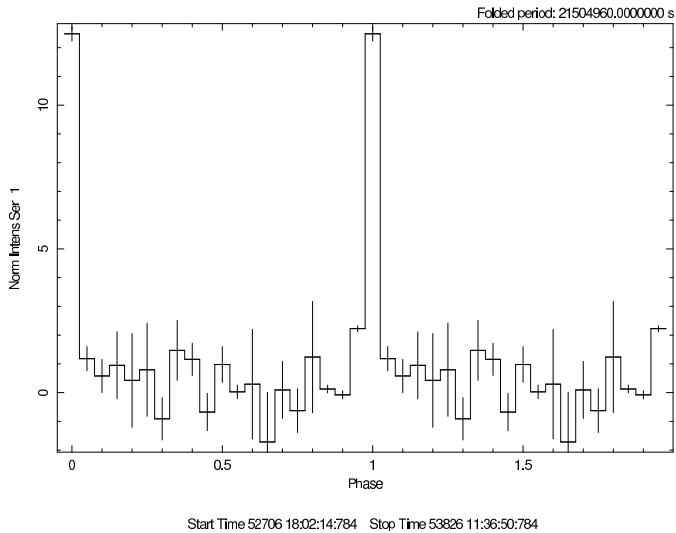}
\caption{All available INTEGRAL data in 18-60 keV range folded at the outburst
period of 248.9d. The phase is the same as in Figure~\ref{fig:fold}.}
\label{fig:intfold}
\end{center}
\end{figure}

A light curve was obtained by analysing all available INTEGRAL data
when the source was within the IBIS/ISGRI field of view (1085 science
windows) and extracting the 18-60 keV flux at the source position in
each science window.  The sampling time for this light curve is
therefore at the typical science window duration of
$\sim$2000s. Folding the light curve on the ephemeris given in Section
6.1 yields the folded light curve shown in
Figure~\ref{fig:intfold}. Though the modulation shown is dominated by
the one major outburst of June 2004 the profile is in good phase
agreement with that shown in Figure~\ref{fig:fold} for the BATSE and
RXTE/ASM outbursts.

Since there are not many data on the second outburst, the results from
the first, previously unreported outburst are presented in this
section.

\begin{figure}\begin{center}
\includegraphics[width=80mm,angle=-0]{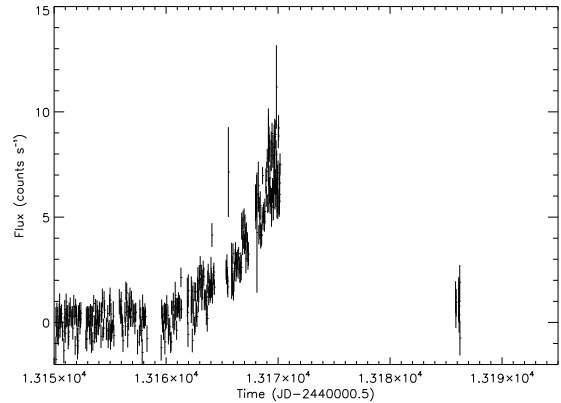}
\caption{INTEGRAL IBIS data showing the outburst of June 2004 in the energy
range 18--60 keV.}
\label{fig:intout}
\end{center}
\end{figure}

A lightcurve of the 2004 outburst is shown in
Figure~\ref{fig:intout} for the photon energy range 18-60 keV. From
this figure it is possible to ascertain that the main burst lasted
less than 23.5 days, and rose in an approximately linear manner with a
doubling time of $\sim$3.5d. In total INTEGRAL observed this outburst
for $\sim$7.9 d.

A 10s binned light curve in the 20--40 keV photon energy range was
obtained from the IBIS/ISGRI data using the \textit{OSA ii\_light}
tool.  The finely binned lightcurve was analysed
using the Lomb-Scargle periodogram method by means of the fast
implementation of Press \& Rybicki (1989).  The resulting power
spectrum shows a clear strong peak at 0.01068 Hz with a power of
$\sim$435.  Using the error formulae derived by Horne \& Baliunas
(1986) we find that this corresponds to a periodicity of
93.6672$\pm$0.0005s.  This period was independently confirmed by the
4--10 keV data from the X-ray camera on-board INTEGRAL,
JEM-X. Furthermore, the period is in very close agreement to that of
93.62$\pm$0.01 reported by Shrader et al. (1999) from the 1993 ASCA
data, indicating very little spin up (or down) over more than a
decade.

\begin{figure}\begin{center}
\includegraphics[width=90mm,angle=-0]{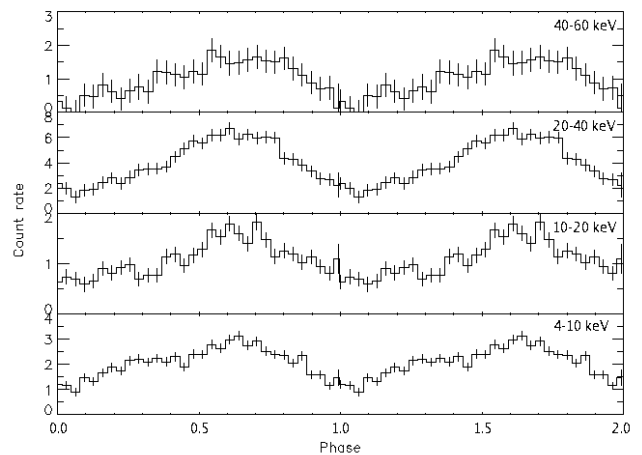}
\caption{INTEGRAL IBIS and JEM-X pulse profiles.}
\label{fig:intpp}
\end{center}
\end{figure}

The phase folded pulse profile was determined in 4 separate energy
channels; 4-10 and 10-20 keV from JEM-X and 20-40 and 40-60 keV from
IBIS/ISGRI. The resulting single peaked profiles are shown in
Figure~\ref{fig:intpp}. From these profiles it was possible to
determine the pulse fractions, ($I_{max} - I_{min})/I_{max}$, to be
72$\pm$8\% in the 4--10 keV band, 70$\pm$10\% in the 10--20 keV band,
80$\pm$10\% in the 40--60 keV, and 100$\pm$30\% in the 40--60 keV
range. The IBIS/ISGRI pulse fractions must be treated with care as they are
derived from the folded fine timing lightcurve and consequently are
not absolutely flux calibrated.

Spectra for the 2004 outburst of GRO J1008--57 were produced using OSA
for both the IBIS/ISGRI and JEM-X instruments.  The source was not
always within the field of JEM-X and consequently there were only 12
science windows of data available spread across the outburst where the
source was detected by that instrument.  The IBIS/ISGRI data for the
same time span comprised of $\sim$80 science windows.  The spectra
were fit simultaneously in
\textit{XSPEC 12.3.0}; the JEM-X data were fit over the 3--20 keV range
and the IBIS/ISGRI data over the 20--100 keV range.  A cross-calibration
constant, C$_{ISGRI}$, was added to all spectral models to account for
any uncertainties in the matching of the two instruments. A systematic
error of 2\% was used.

\begin{table}
\centering
\caption{The results of the simultaneous spectral fits of the JEM-X and ISGRI data.}

\begin{tabular}{|c|cc}
\hline
Parameter	&	Cut-off Power Law	&Thermal Bremsstrahlung \\
\hline
C$_{ISGRI}$	&		1.6 $\pm$ 0.1	&	1.5 $\pm$ 0.1 \\
cut-off energy	&		8.0 $\pm$ 1.0 keV	&	-	 \\
Fold energy	&		21 $\pm$ 2 keV	&	-		 \\
Photon index	&		1.4 $\pm$ 0.1	&		-	 \\
kT 		&	-			&	21 $\pm$ 1 keV \\
$\chi$$^{2}$/dof&		150.46/146	&	162.31/148	\\								
\end{tabular}

\label{tab:spectra}
\end{table}

\begin{figure}\begin{center}
\includegraphics[width=60mm,angle=-90]{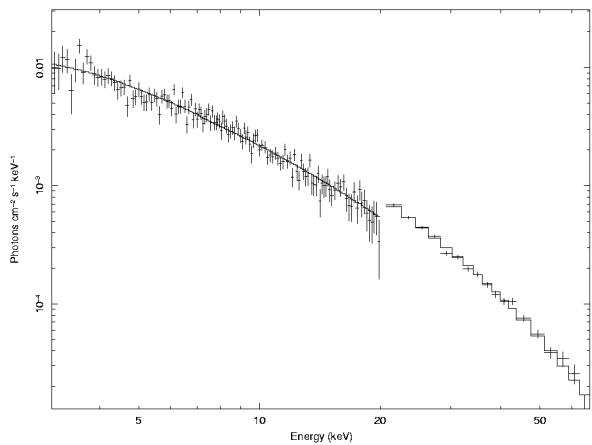}
\caption{INTEGRAL combined IBIS and JEM-X spectrum obtained by
deconvolving the data using the Cut-off Power Law model.}
\label{fig:intspec}
\end{center}
\end{figure}

Two spectral models were fit to the data, a power-law with a high
 energy cut-off and a thermal bremsstrahlung model.  Both models fit
 equally well and have a reduced chi-square of $\sim$1; the details of
 the fit parameters are shown in Table~\ref{tab:spectra}.  An
 attempt was made to introduce an absorption component to both,
 however the fit could not resonably constrain the parameter.
 Consequently, an absorption component fixed to the anticipated
 galactic absorbing column density of $\sim$1.5 $\times$ 10$^{22}$
 cm$^{-2}$ was added.  The unfolded spectra with the high-energy
 cut-off power law model fitted is shown in Figure~\ref{fig:intspec}.

The fluxes
given by the spectral models were 5.7 $\times$ 10$^{-10}$ erg
cm$^{-2}$ s$^{-1}$ in the 3--20 keV band and 4.1 $\times$ 10$^{-10}$
erg cm$^{-2}$ s$^{-1}$ in the 20--100 keV band. Taking these numbers
combined with the source distance of 5kpc (Coe et al, 1994) produces
a flux estimate of (1--2) $\times$ 10$^{36}$ erg s$^{-1}$. This value
is in good agreement with the level of X-ray luminosity estimated from
the small accretion-torque driven pulse period changes reported in Section 4.

\section{Discussion}

\subsection{X-ray outburst ephemeris}

Using the clear outburst profile presented in Figure~\ref{fig:fold}
for the period 2001-2006 it is possible to determine an accurate
ephemeris. We assume the binary period is 248.9d (Levine \& Corbet,
2006) and chose the highest bin in the figure to be defined as phase
0.0. From this the following outburst ephemeris is determined:

$T_{outburst}$ = 50186$\pm$4 + 248.9n

\noindent where the dates are in Modified Julian Day format and n is
the outburst number.

If the time of outburst is compared to the predicted time of periastron
determined from the pulse period variations
(see Table 2), it agrees within 1.5d for the MJD 49189 outburst,
indicating no significant phase shift between these two events.

If we use this RXTE/ASM ephemeris then we can determine that the phase
of the two INTEGRAL outbursts discussed here are 0.99 and
0.00. Furthermore, the phase of the peak of the 1993 BATSE outburst
reported in Shrader et al. (1999) is 0.01. Thus all three reported
events fit very comfortably with the above ephemeris and hence are
almost certainly Type I outbursts. This is supported by comparing the
phase of the peaks of outbursts as seen in the folded lightcurves from
BATSE, RXTE/ASM and INTEGRAL all of which seem to match very
well. There is no evidence for the phase changes in outburst seen in
EXO 2030+375 and explained by Wilson et al. (2002) as related to global
one arm oscillations in the circumstellar disk. Though there are
changes in the disk, evidenced by the overall shape changes of the
H$\alpha$ profile seen in Figure~\ref{fig:ha}, they are obviously not
great enough to result in such outburst phase shifting.

\subsection{Determining the circumstellar disk size and its
relationship to X-ray activity.}

Figure~\ref{fig:ew2} shows the H$\alpha$ equivalent width
measurements over the last 13 years.  
Although the H$\alpha$ equivalent width is not a
direct measurement of the size of the circumstellar disk the data in
this figure can be
used as an indicator of growth and decline in the disk. To this effect
we interpret the significant decline in the H$\alpha$ flux, followed
by a strong recovery as a period of disc shrinking and regrowth.  This
idea is consistent with the lack of observed X-ray outbursts of
GRO~J1008$-$57 during 1996--2001 - see Figure~\ref{fig:fold}.

Another feature to take note of in Figure~\ref{fig:ew2} is the maximum
measured H$\alpha$ equivalent width.  Reig, Fabregat \& Coe (1997)
show that this maximum H$\alpha$ equivalent width is correlated with
the orbital period in a Be/X-ray binary and the observed correlation
can be explained by a viscous circumstellar disk truncated by the
resonant torque of the orbiting neutron star (Negueruela et al.,
2001).  The system under discussion here spends a large fraction of
the observed timescale at an H$\alpha$ equivalent width of close to
$-20$\AA, while undergoing Type~I X-ray outbursts.  This may suggest
that the $-20$\AA ~EW measurement describes the circumstellar disk at
a size that puts it close to the $L_1$ point at periastron.

To estimate the size of the H$\alpha$ emitting region of the
circumstellar disk we make used of the peak separation measurements in
Table~1 and the relation from Huang (1972).
 



where $r$ is the disk radius, $i$ is the inclination and $\Delta V$ is
the peak separation. The average of the three measurements taken while
the system was in normal outburst was determined to be 236$\pm$15 km/s
and used in the following determination of the disk size.

From the blue spectrum we determine the spectral type of the optical
counterpart of GRO~J1008$-$57 to be B1 - B2 Ve.  From Allen (1973) we
estimate the mass and radius of such a star to be 15\,M$_{\odot}$ and
7\,R$_{\odot}$ respectively.  Then, using the measured values of
$a_x\sin i$ and assuming that the Be star disk is in the plane of the
orbit we find an inclination angle of $i=36^{\circ}$ for the
circumstellar disk of GRO~J1008$-$57.  These stellar parameters give
an estimated H$\alpha$ disk size of 72\,R$_{\odot}$ and a ratio of
disk size to L$_{1}$ point (measured from the centre of the Be star)
of R$_{d}$/R$_{L}$ $\sim$ 0.64

In comparison, if the data for 1997 are used with a $\Delta$V of 350
km/s, then this results in a much smaller disk of the order of
33\,R$_{\odot}$ in size.

Even the 72\,R$_{\odot}$ is much smaller than the truncation radius
predicted by the viscous disk theory for this Be/X-ray binary  by Okazaki
\& Negueruela (2001).  Using slightly different stellar and orbital
parameters those authors predict that the circumstellar disk of GRO
J1008-57 is most likely truncated at the 7:1 or 8:1 resonance
radius. At such a radius the disk verges on the L$_{1}$ point at
periastron, thus explaining the frequently observed Type I outbursts. 

Exploring the upper and lower limits of the disk size estimates
presented in this work requires the identification of the major
uncertainties affecting the results.The uncertainty in the spectral
types introduces a range of possible masses for the Be star of 10 --
18 M$_{\odot}$. Using extreme upper limits for both the stellar mass
and $a_x\sin i$, with the lower limit on $\Delta$V, we find that
R$_{d}$/R$_{L}$ $\sim$ 0.86. At the other end of the range we find  
R$_{d}$/R$_{L}$ $\sim$ 0.51

However, it is worth bearing in mind that, contrary to the assumption
in most disk models, the Be star disk is probably not isothermal.
Thus the physical size of the disk is not constrained to the H$\alpha$
emitted size of the disk. For example, in the case of the isolated Be
star $\Psi$ Per the H$\alpha$ disk size has been measured by direct
interferometry to be around ten stellar radii (Quirrenbach et al.,
1997), whereas similar radio observations (Dougherty \& Taylor 1992)
suggest that material may be present as far away as hundreds and even
thousands of stellar radii. So, in the case of GRO J1008--57 the
totality of the disk may extend beyond the radius measured here from
the H$\alpha$ data. This is indirectly confirmed by the very fact that
we see regular X-ray outbursts, suggesting the disk material must,
indeed, reach the L$_{1}$ point.

\section{Conclusions}

The study of pulse period changes in GRO J1008--57 from the BATSE data
have allowed a precise orbital determination. The orbital period found
agrees, within errors, with that determined by Levine \& Corbet (2006)
from X-ray outburst cycles. These outbursts seem to be very reliable
in nature, showing no variation with X-ray energy, nor are they
correlated with the changes in the H$\alpha$ profile, as was seen in
the case of EXO 2030+375 (Wilson et al, 2002).  However, the
substantial changes in the H$\alpha$ EW correlate nicely with the
existence and changing levels of the X-ray outburst sizes.  In
addition, using the peak separation measured from the H$\alpha$
profiles an indication was provided of the disk size which is
in general agreement with the detailed models of Okazaki \& Negueruela
(2001). The results are a direct confirmation of the value of
multiwavelength observations of these complex systems.

\section{Acknowledgements}

This paper uses observations made from the South African Astronomical
Observatory (SAAO), and X-ray data provided by the ASM/RXTE team.
VAM acknowledges support from the South African NRF and the British Council
in the form of a SALT/Stobie studentship.

\bsp

\label{lastpage}

\end{document}